\documentclass[useAMS,usenatbib]{mn2e}
\usepackage{graphicx}
\usepackage{amsmath}
\usepackage{color}
\usepackage{amssymb}

\title[Clustering of Gamma-ray Bursts]
{Investigation of Redshift- and Duration-Dependent Clustering of Gamma-ray Bursts}
\author[T. N. Ukwatta and P. R. Wo\'zniak]{
T. N. Ukwatta$^{1}$\thanks{E-mail: tilan@lanl.gov; tilan.ukwatta@gmail.com} and P. R. Wo\'zniak$^{2}$\\
$^{1}$Director's Postdoctoral Fellow, Space and Remote Sensing (ISR-2), Los Alamos National Laboratory, Los Alamos, NM 87544, USA.\\
$^{2}$Space and Remote Sensing (ISR-2), Los Alamos National Laboratory, Los Alamos, NM 87544, USA.}

\begin{document}



\maketitle

\label{firstpage}

\begin{abstract}
Gamma-ray bursts (GRBs) are detectable out to very large distances and as such
are potentially powerful cosmological probes. Historically, the angular distribution
of GRBs provided important information about their origin and physical properties.
As a general population, GRBs are distributed isotropically across the sky. However,
there are published reports that once binned by duration or redshift, GRBs display
significant clustering. We have studied the redshift- and duration-dependent
clustering of GRBs using proximity measures and kernel density estimation.
Utilizing bursts detected by BATSE, $Fermi$/GBM and $Swift$/BAT,
we found marginal evidence for clustering in very short duration GRBs lasting
less than 100 ms. Our analysis provides little evidence for significant redshift-dependent
clustering of GRBs.
\end{abstract}


\section{Introduction} \label{introduction}

Gamma-ray bursts (GRBs) are often referred to as the biggest explosions in the Universe
since the Big Bang. Their powerful prompt electromagnetic emission and afterglows make them
detectable out to very high redshifts $z > 10$ \citep{lamb2000}. Therefore, GRBs offer
a potential probe to study inhomogeneities and anisotropies in the Universe on the largest
scales.

Historically, the angular distribution of GRBs provided important information about their
origin and physics. These early studies have shown that the sky distribution
of GRBs is isotropic \citep{Meegan1992, Briggs1996, Tegmark1996}, providing early indications---before
the discovery of afterglows---that GRBs are at cosmological distances. While the isotropic
sky distribution is well established for long duration GRBs, there have been reports
of clustering for short
GRBs (T90 $<$ 2 s)~\citep{Balazs1998, Magliocchetti2003}, very short
GRBs (T90 $<$ 100 ms, VSGRB)~\citep{Cline1999,Cline2005, Cline2011}, and intermediate
duration GRBs (2 s $<$ T90 $<$ 8 s)~\citep{Meszaros2000, Litvin2001}. 
Other reports of clustering are based on parameters
such as spectral lags~\citep{Ukwatta2010, Ukwatta2012}. For example, long-lag
GRBs tend to cluster in the supergalactic plane~\citep{Norris2002,Foley2008}.
If confirmed, these observations may point to the existence of new 
sub-populations of GRBs, possibly with distinct progenitors.

Thanks to the rapid localizations delivered by the $Swift$ mission \citep{Gehrels2004}
we now have more than 300 GRBs with redshift measurements. This opens up the exciting
possibility to explore the distant universe using high-redshift GRBs. Recently,
\cite{Horvath2012} have reported evidence of strong anisotropy in the observed sky distribution
of GRBs in the redshift range between 1.6 and 2.1 based on the distribution of the angular distance
to the n-th nearest neigbour and a two-dimensional Kolmogorov-Smirnov (KS) test.
Detection of very large scale structures such as the one implied by the \cite{Horvath2012} study
would have profound implications for the cosmological principle which states that on average
the Universe is homogeneous and isotropic.

Here we investigate the angular sky distribution of GRB sub-populations
using various density and proximity estimators. Our study is divided into two parts.
First, we investigate redshift-dependent GRB clustering using a sample of 311 $Swift$
bursts with measured redshifts. We use full Monte Carlo simulations to generate the relevant
probability distributions and evaluate the significance of potential angular structures.
This analysis is limited to bursts detected by $Swift$ because an overwhelming majority of GRBs
with redshift measurements are in that category and also because of the possibility to correct
the density estimates to account for irregularities in the exposure time allocated to various
parts of the sky. In the second part of the paper we use combined samples of GRBs detected
by multiple instruments to study duration-dependent angular distributions. To examine
the significance of clustering results we construct approximate exposure maps for
multi-instrument samples assuming that the intrinsic distribution of long GRBs is isotropic.

The paper is organized as follows. In Section~\ref{methodology} we describe our methodology.
In Sections~\ref{z_clustering} and ~\ref{t90_clustering} we present the analysis of the
redshift-dependent and duration-dependent GRB clustering.
In Section~\ref{discussion} we discuss the caveats and implications of our findings,
and Section~\ref{conclusions} summarizes the conclusions.

\section{Density Estimation} \label{methodology}

The search for clustering in the angular distribution of GRBs begins with computing all-sky
maps for a number of density and proximity measures. Detailed Monte Carlo simulations are
then used to establish the statistical significance of the observed clumps and investigate
possible systematics.

\subsection{N-th Nearest Neighbor Density Estimator}

The distance to the n-th nearest neighbor is a widely used proximity measure
in astronomy~\citep{Ivezic2014}. In the case of sky distributions this distance
is the angular distance between two distinct points on the sphere. One can obtain
a two-dimensional density measure by calculating the area enclosed within some radius
and then inverting it.
Let $\theta_i$ be the angular distance to the i-th nearest neighbor. The area $a_i$
enclosed by $\theta_i$ is
\begin{equation}\label{eq:vol}
a_i = 2 \pi (1 - \cos \theta_i).
\end{equation}
For $n > 2$, we can write an unbiased n-th nearest neighbor density estimator $\rho_n$
and its variance $\sigma^2$ as
\begin{equation}\label{eq:nnnd}
\rho_n = \frac{n-1}{a_n}, ~~~~~~~~~~~~~
\sigma^2({\rho_n}) = \frac{\rho_n^2}{n-2}.
\end{equation}
In the uniform density case, $\rho_n$ is a sufficient statistic, meaning that all
neighbors closer than the n-th do not contribute additional information.
For details see the discussion in~\citep{Wozniak2012}.

\subsection{Gaussian Kernel Density Estimator}

The choice of $n$ for the nearest neighbor density estimator is somewhat arbitrary
and reflects a tradeoff between the variance and the spatial scale of the observed density
fluctuations. A better approach is to use kernel density estimators, a class of
non-parametric estimators with a flexible functional form that can utilize all available
data. A kernel density estimator on the sphere is defined as
\begin{equation}\label{eq:kernel_estimate}
\hat{f}_{h}(x) = \frac{1}{N \Omega(h)} \sum_{i=1}^{N} K \bigg(\frac{\theta(x, x_i)}{h}\bigg).
\end{equation}
Here $x$ is the location at which the density is predicted based on the set of $N$
data points $x_i$. The smoothing length $h$ is sometimes referred to as the bandwidth.
The kernel function is normalized so that $\hat{f}_{h}(x)$ represents a probability density
over the spherical measure of area $d\Omega = \sin \theta d\theta d\phi$:
\begin{equation}\label{eq:kernel_int}
\Omega(h) = \int K\bigg({\theta \over h}\bigg)d\Omega(\theta).
\end{equation}
To obtain the number density of GRBs per unit solid angle for a particular sample,
Equation~\ref{eq:kernel_estimate} must be multiplied by $N$, the total number of
bursts in the sample covering the entire sky. A variety of kernel functions have
been explored in the literature~\citep{Klemela2000}. For this study we choose the
Gaussian kernel function defined as
\begin{equation}\label{eq:gussian_kernel}
K(y) = \exp \bigg(-{1 \over 2}y^2\bigg).
\end{equation}
Our density estimator is therefore
\begin{equation}\label{eq:gussian_kernel_estimator}
\hat{f}_h(x) = \frac{1}{N \Omega(h)} \sum_{i=1}^{N} \exp\bigg(\frac{-\theta_i^2}{2 h^2}\bigg),
\end{equation}
where
\begin{equation}\label{eq:gussian_kernel_norm}
\Omega(h) = 2 \pi\int_{0}^{\pi} \exp\bigg(\frac{-\theta^2}{2 h^2}\bigg) \sin\theta d\theta.
\end{equation}
Here $\theta_i = \theta(x, x_i)$ is the spherical distance between an arbitrary line of
sight $x$ and the location $x_i$ of the i-th GRB in the sample. Each term in $\hat{f}_{h}$
is shown to have the same norm $\Omega(h)$ by shifting the pole of the spherical
coordinate system $(\phi, \theta)$ to the corresponding data point.

The performance of kernel density estimators does not depend strongly on the choice of
the kernel function, however it is crucial to select a good value of the
smoothing length. The bandwidth $h$ can be robustly optimized by minimizing the mean
integrated squared error (MISE; e.g. ~\cite{Ivezic2014}).
\begin{equation}\label{eq:optimum_h_estimator}
\int(\hat{f}_{h}-f)^2 d\Omega \\
= \int\hat{f}_{h}^2 d\Omega - {2 \over N} \sum_{i=1}^{N}\hat{f}_{h, -i}(\theta_i) + \int f^2 d\Omega
\end{equation}
The last term that includes the unknown true density $f(x)$ does not depend on
$h$ and is effectively a constant offset. The middle term is proportional to the
expectation value of the density estimator. It is approximated here using leave-one-out
cross-validation, i.e. by taking a mean of $N$ independent density estimates at the
location of each burst $i$ based on $N-1$ remaining bursts. Finally, the first term is
evaluated by the brute force numerical integration because
the order of the summation and integration can no longer be switched.

\section{Redshift-Dependent Clustering of GRBs}\label{z_clustering}

\subsection{GRB Sample}

\begin{figure}
\includegraphics[width=84mm]{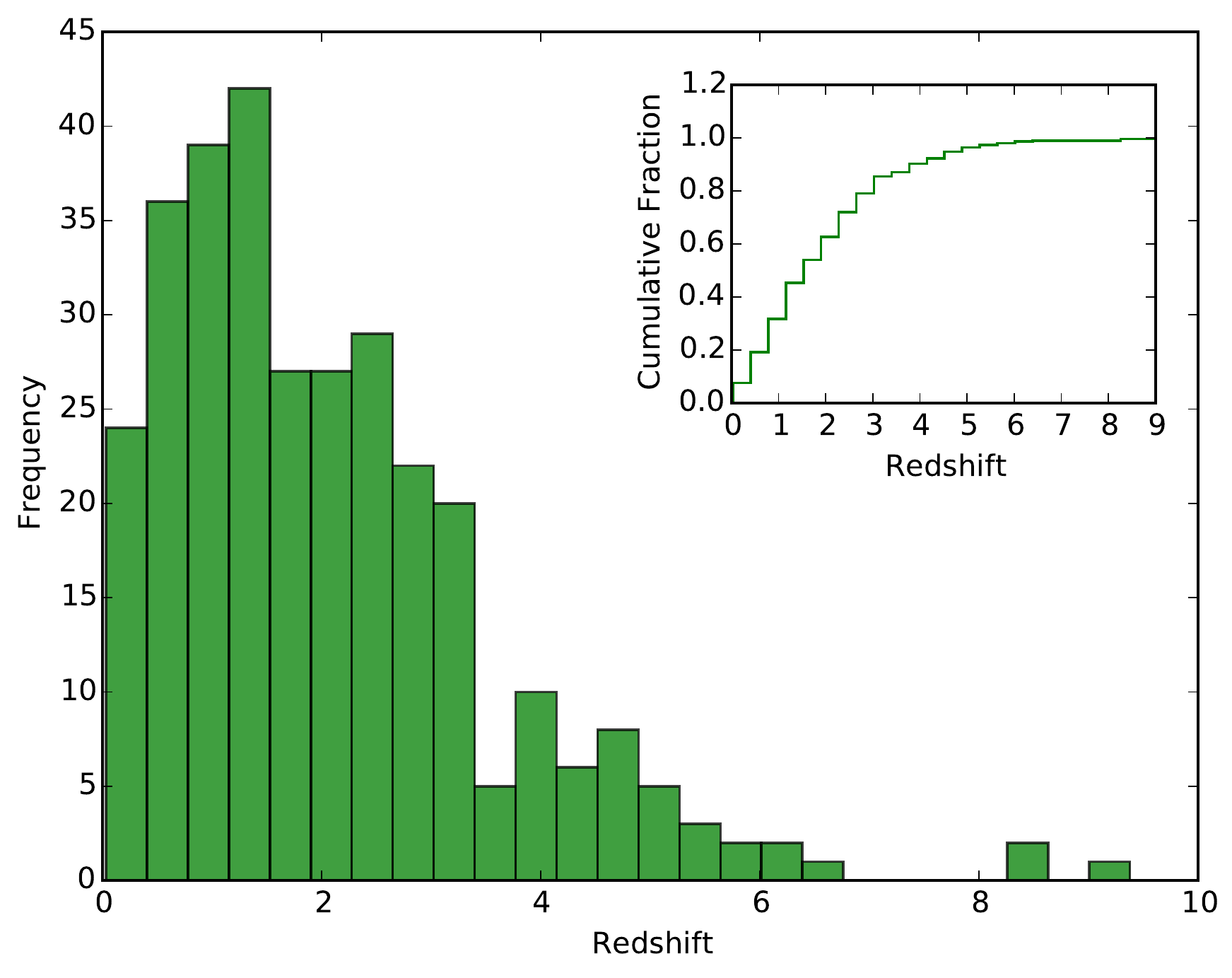}
\caption{Redshift distribution for a sample of 311 $Swift$ bursts.}\label{zhisto}
\end{figure}

Clustering in the sky distribution of GRBs over a limited range of redshift indicates
the presence of large scale structures in the Universe. However, statistically significant
overdensities may also occur due to observational biases that are unrelated
to any physical groups. In order to construct a uniform, high-quality sample, we limit
this part of our study to GRBs detected by $Swift$. Since most bursts with a known redshift
are in fact detected by $Swift$, this selection does not drastically reduce
the sample size. Our sample includes 311 $Swift$ GRBs with redshift measurements from the
third $Swift$ Burst Alert Telescope Gamma-ray Burst Catalog~\citep{Lien2015}.
The redshift distribution of our GRB sample is shown in Figure~\ref{zhisto}.
Both short and long GRBs are included in the sample. All GRBs are believed to be tracers of
galaxies and matter, so in principle no further selection cuts are needed to search
for clusters. The sample is strongly skewed toward low redshift. There are
50 bursts with $z>3.5$ and only 25 with $z>4$.

\subsection{GRB Density Map}

\begin{figure}
\includegraphics[width=84mm]{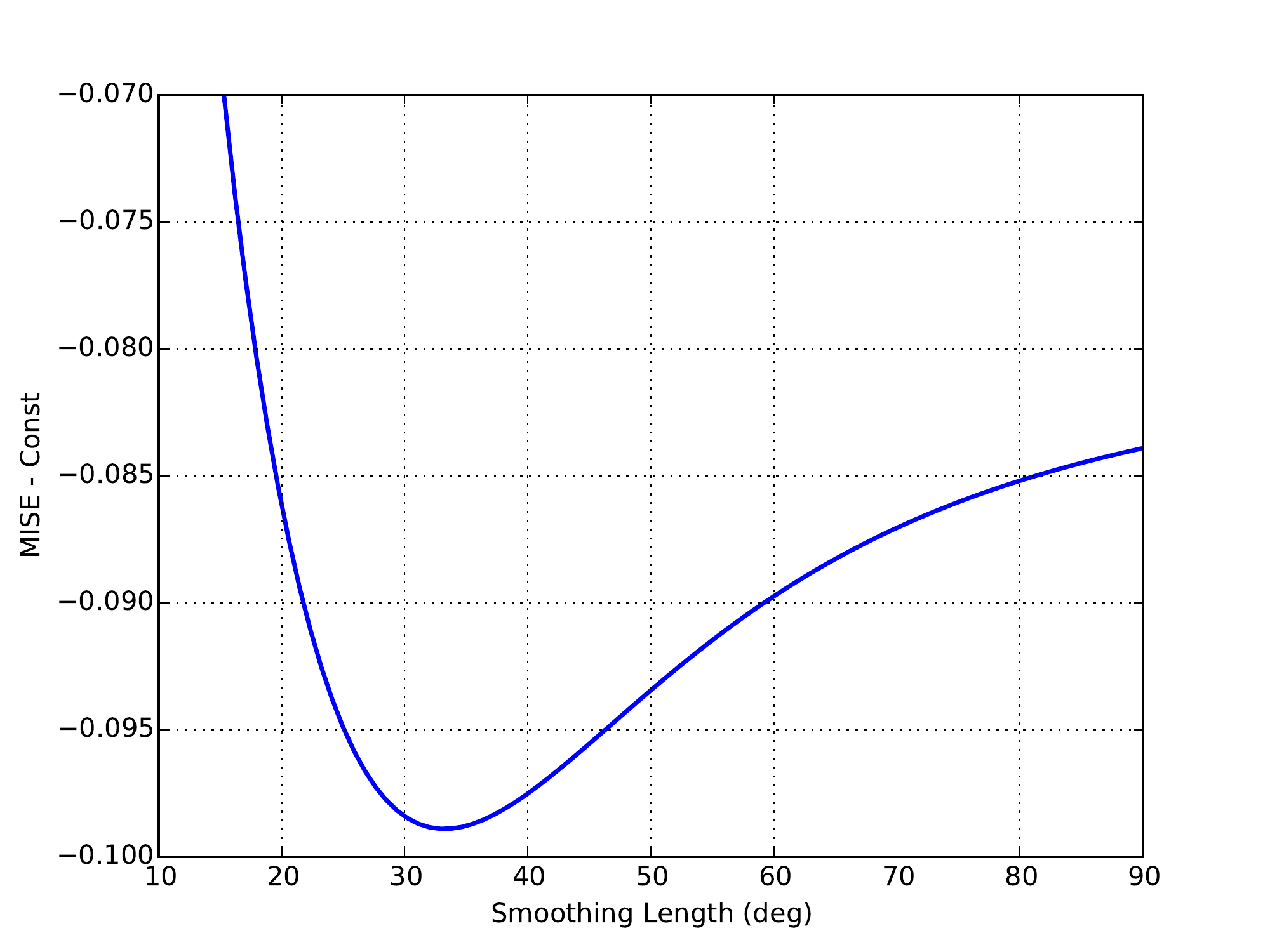}
\caption{Mean integrated square error (MISE) as a function of the smoothing length
for a subsample of 34 bursts from Figure~\ref{z_density_map}. The optimal smoothing
length is $\sim33$ deg found by minimizing MISE in Equation~\ref{eq:optimum_h_estimator}.}\label{mise}.
\end{figure}

\begin{figure*}
\includegraphics[width=0.9\textwidth]{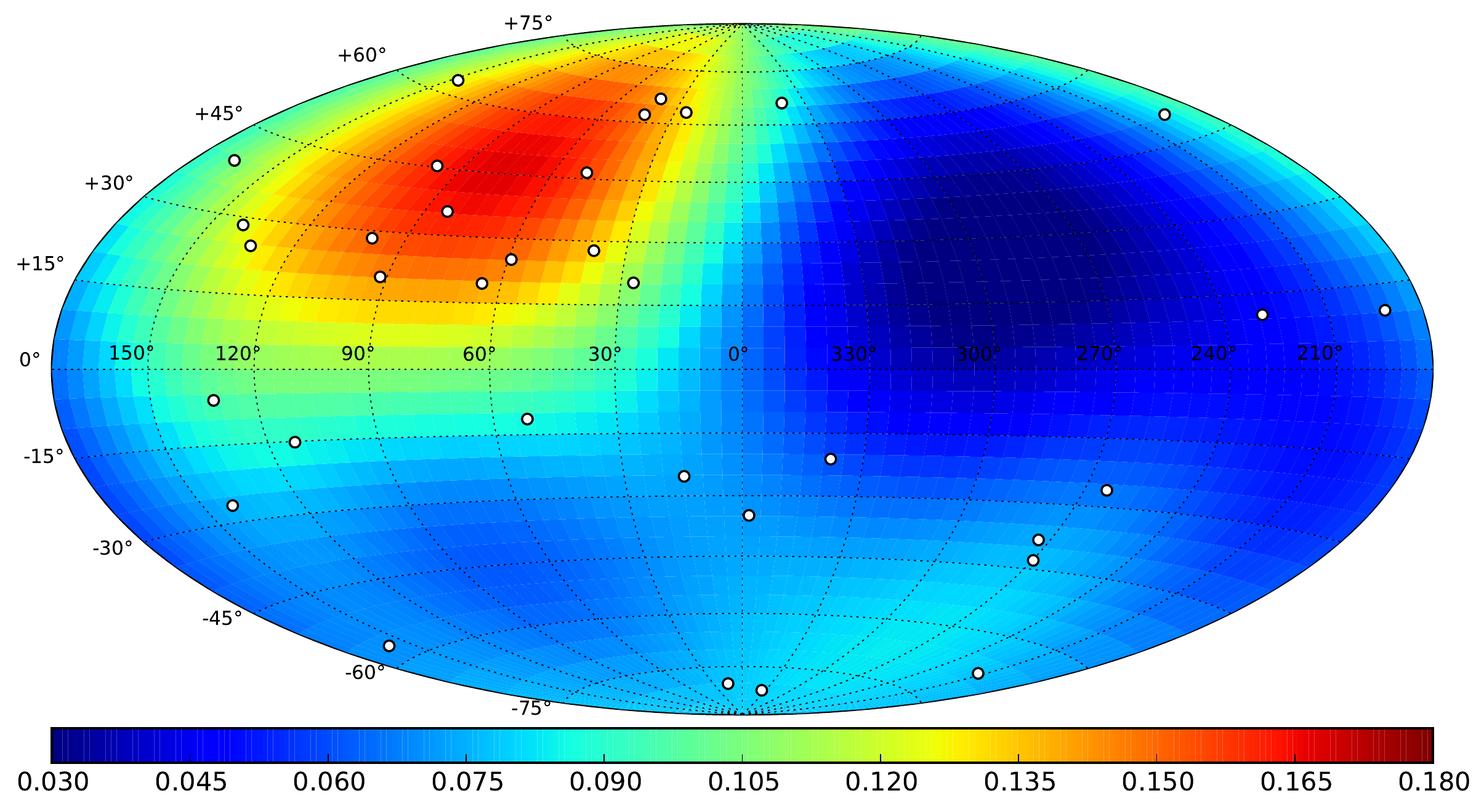}
\caption{Density map in galactic coordinates for a subsample of 34 GRBs
in the redshift range 1.6 $< z <$ 2.1. The color-coded quantity is the probability
density function (PDF) normalized to the full sky. The optimal smoothing length is 33 degrees.
Circles indicate the actual burst locations. The maximum and the minimum density values
in this map are 0.167 and 0.026, correspondingly.
The probability of generating this density contrast by chance estimated using a Monte Carlo
simulation is 0.013 assuming that the true sky distribution is uniform, 0.022
assuming that burst detections follow the $Swift$ exposure map, and 0.025 considering
both $Swift$ exposure function and redshift selection bias.
}\label{z_density_map}
\end{figure*}


The first step in our search for GRB clustering is computing all sky density maps for various
samples and redshift ranges using Equation~\ref{eq:gussian_kernel_estimator}. Throughout
this paper each map is normalized to represent a probability density function (PDF)
that integrates to $1$ over the entire sphere ($4\pi$ solid angle). The number density
of bursts per steradian depends on the sample and is easily obtained by multiplying the PDF
by the total number of data points in the sample. Other quantities such as the total
exposure time per line of sight are also best visualized as PDFs and scaled as needed.

\cite{Horvath2012} examined the angular distribution of GRBs in multiple redshift bins
using a two-dimensional KS test and the n-th nearest neighbor distance. Based on a somewhat
different data set they found that the apparent anisotropy is strongest for $1.6 < z < 2.1$.
In our GRB catalog this redshift range contains 34 bursts and also approximately
maximizes the observed density contrast. To enable a direct comparison with \cite{Horvath2012}
we focus our analysis of redshift-dependent clustering in the same redshift bin.
The optimal smoothing length $h$ is calculated by minimizing MISE (Equation~\ref{eq:optimum_h_estimator}) as shown in Figure~\ref{mise} for $Swift$ bursts with $1.6 < z < 2.1$.
The best value of $h$ depends on both the size of the sample and the angular scale
of the most significant clusters. In this particular case $h\sim33$ deg.
The resulting GRB density map is shown in Figure~\ref{z_density_map}.
Taken at face value, the map may suggest that the distribution is clumpy,
a rigorous estimate of the significance of the observed density fluctuations
is required to draw any conclusions, especially for small samples.

\begin{figure}
\includegraphics[width=84mm]{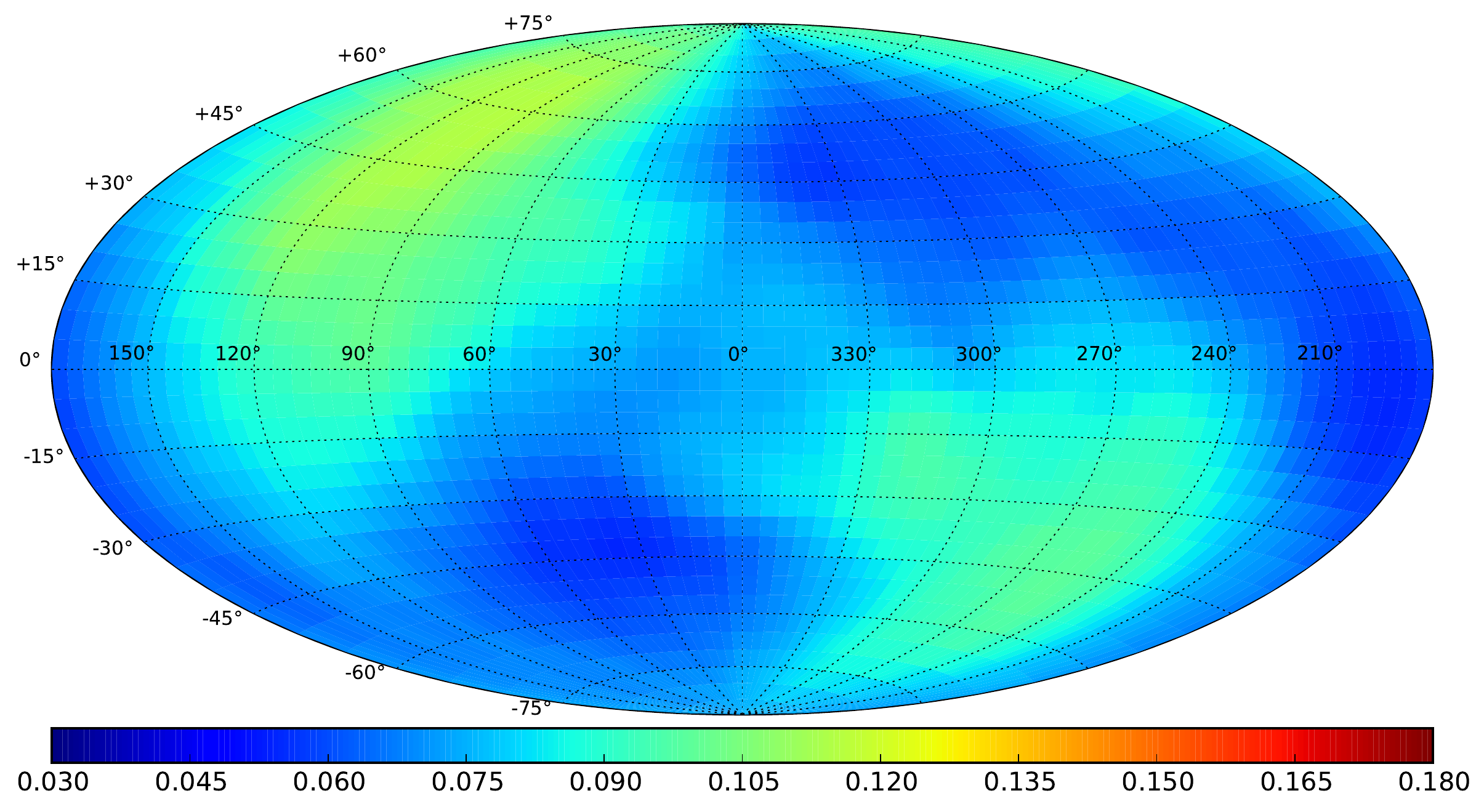}
\caption{$Swift$/BAT exposure map in galactic coordinates from 104 months
of observing. The partial covering fraction is set to 100\% over the entire
BAT field of view~\citep{Baumgartner2015}. The color scale indicates the probability
density function of observing a particular line of sight. To obtain the total exposure
time in seconds the PDF should be multiplied by $3.41 \times 10^8$.}\label{expmap}
\end{figure}

Variations in the total observing time from one line of sight to another
may introduce spurious density fluctuations in GRB samples. Fortunately, this
information is available for $Swift$ in the relevant time interval.
Figure~\ref{expmap} shows the $Swift$ exposure map in galactic coordinates
based on 104 months of observing. The partial covering fraction was set to
100\% over the entire BAT field of view~\citep{Baumgartner2015}. Similar to
the number density plots, the density and the corresponding color scale were converted
to a probability density for easy comparisons. Multiplying by $3.41 \times 10^8$ converts back to the total
exposure time in seconds. It is intriguing that both the density map (Figure~\ref{z_density_map})
and the exposure map (Figure~\ref{expmap}) display similar peaks and valleys
in roughly the same directions, which points to a possible observational bias.
However, Figure~\ref{z_density_map} shows a peak-to-valley density contrast around
$\sim6.5$, while the exposure time ratio between the most and the least
observed areas in Figure~\ref{expmap} is only a factor of two or so.

\begin{figure}
\includegraphics[width=84mm]{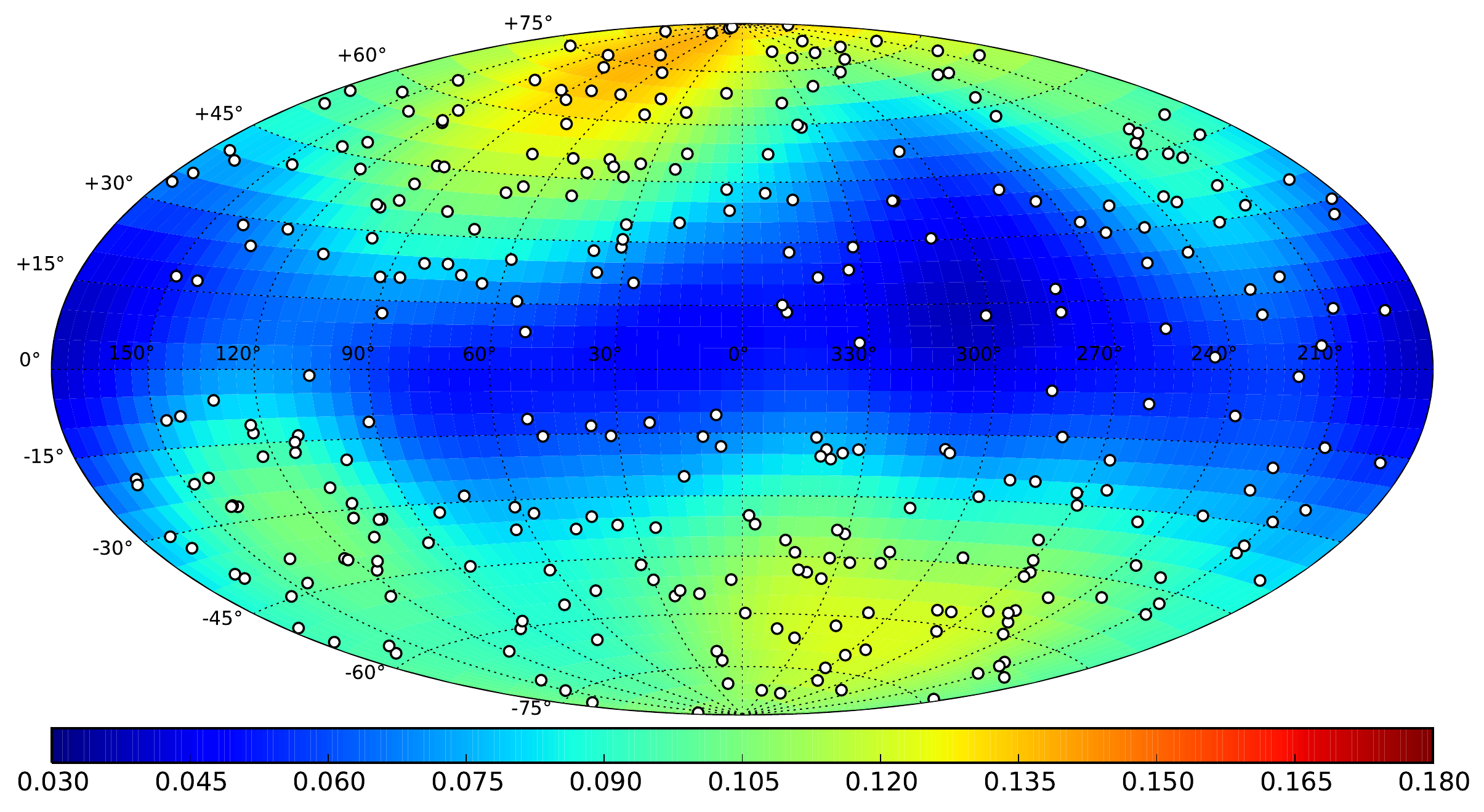}
\caption{Density map in galactic coordinates for a sample of 311 GRBs
with measured redshifts. The optimal smoothing length is 20 degrees.
Circles indicate the actual burst locations. The maximum and the minimum density values
in this map are 0.139 and 0.038.}\label{expmap_redshift}
\end{figure}

\begin{figure}
\includegraphics[width=84mm]{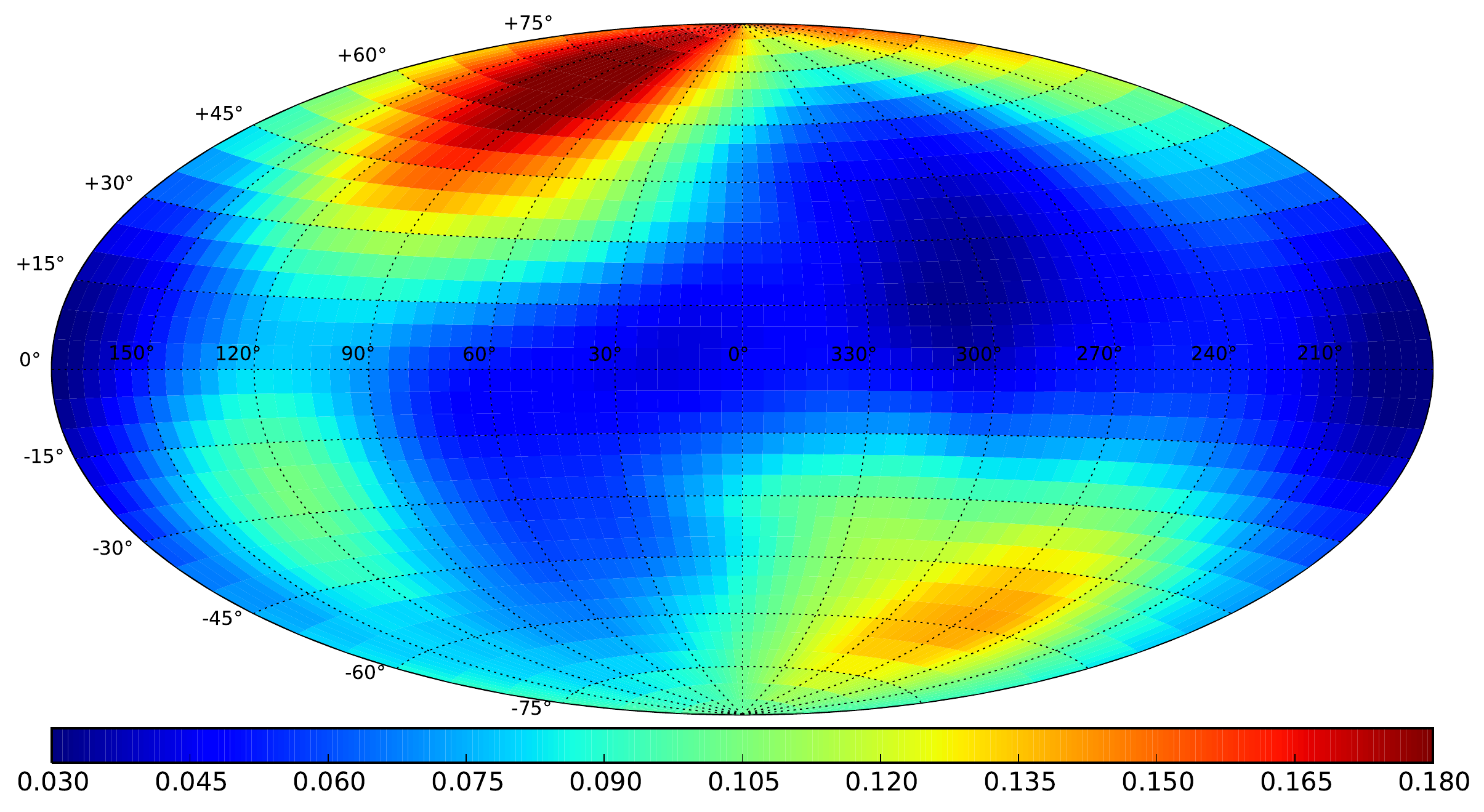}
\caption{Combined probability density map that includes both
the $Swift/BAT$ exposure map and the bias due to the required redshift measurement.
The maximum and the minimum density values
in this map are 0.188 and 0.028.}\label{expmap_combined}
\end{figure}

Another important effect to consider is the selection bias
introduced by redshift measurements. Since it is difficult to measure redshift of a GRB
near the galactic plane, we expect to find GRBs with redshift measurements preferentially
at high galactic latitudes. Indeed, the effect is clearly visible once all redshift bins
are combined for better statistics (Figure~\ref{expmap_redshift}). In order to account for both the $Swift$
exposure and redshift selection, we need to use a combined $Swift$ probability
map shown in Figure~\ref{expmap_combined} obtained by multiplying maps shown in
Figure~\ref{expmap} and Figure~\ref{expmap_redshift}. This combined density map
is somewhat similar to the GRB distribution in Figure~\ref{z_density_map}
with about the same peak-to-valley density contrast $\sim$ 6.7.

We evaluate the significance of the observed density contrast using a Monte Carlo
simulation. The simulator generates 5000 synthetic samples of 34 GRBs with random locations
following three different scenarios for the underlying distribution:
1) the $Swift$ exposure map from Figure~\ref{expmap},
2) the combined probability map from Figure~\ref{expmap_combined} and
3) the uniform PDF over the entire sphere.
A density map similar to Figure~\ref{z_density_map} and the maximum density value
is then computed for each sample. The distribution of these maxima is then used to derive
the probability that the maximum density observed in actual data
(Figure~\ref{z_density_map}) may occur due to random fluctuations referred to as the p-value.
The probabilty of getting the maximum density value seen in
Figure~\ref{z_density_map} from random fluctuations following a uniform distribution
is 0.013 or 1.3\%. The same density peak is easier to generate by chance
from the $Swift$ exposure map and the corresponding p-value is 0.022.
It is even easier to generate (p-value of 0.025) with the combined map which also incorporates
redshift selection effects.

\begin{figure}
\includegraphics[width=84mm]{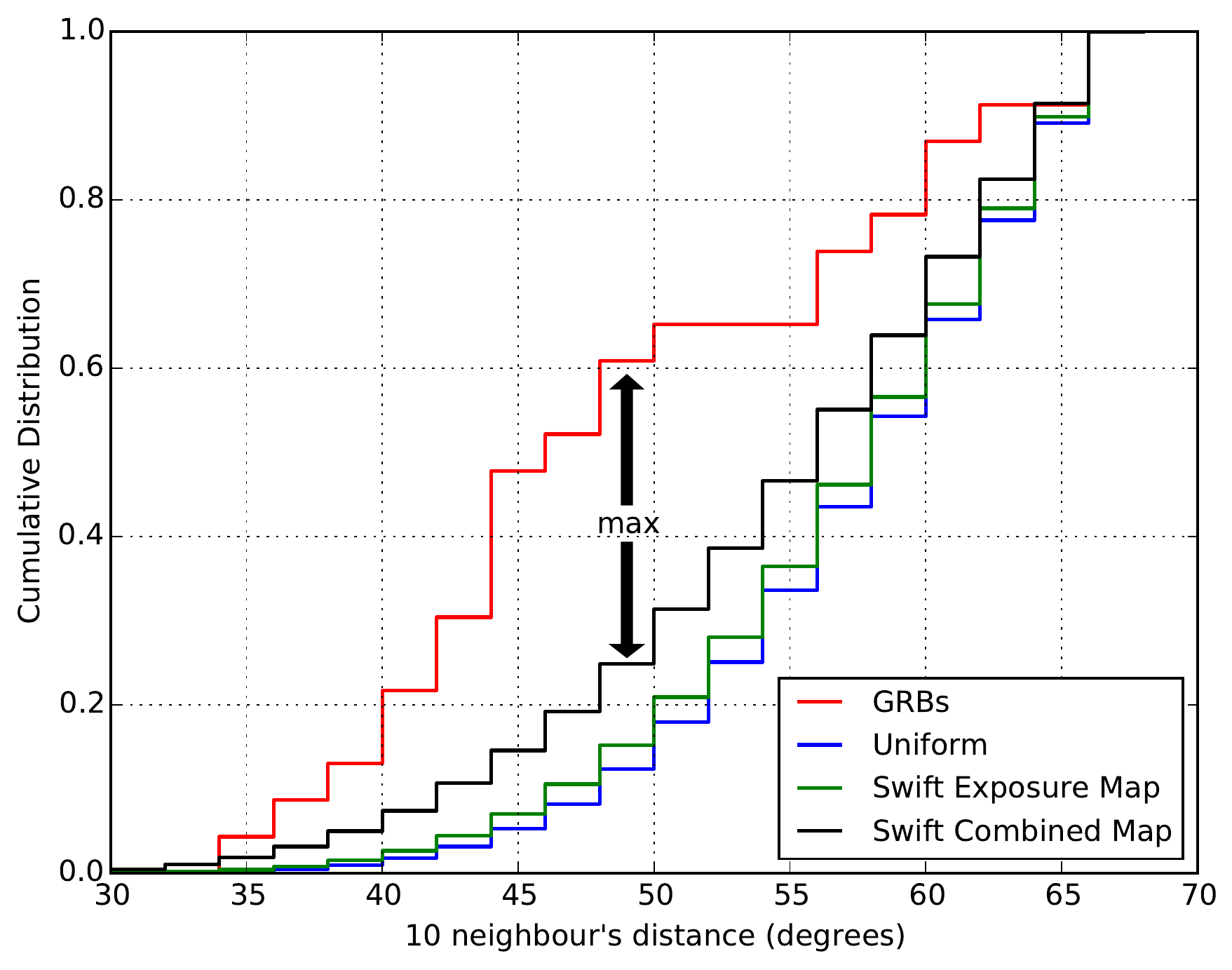}
\caption{Observed 10-th nearest neighbour distance distribution compared to the mean
of 5000 GRB samples generated from the $Swift$ exposure map.
The p-value is 0.016 for the deviation from the uniform distribution,
0.022 for the deviation from the $Swift$ exposure function, and 
0.127 for the deviation from the combined $Swift$ probability map.
}\label{nnn_redshift}
\end{figure}

\begin{figure}
\includegraphics[width=84mm]{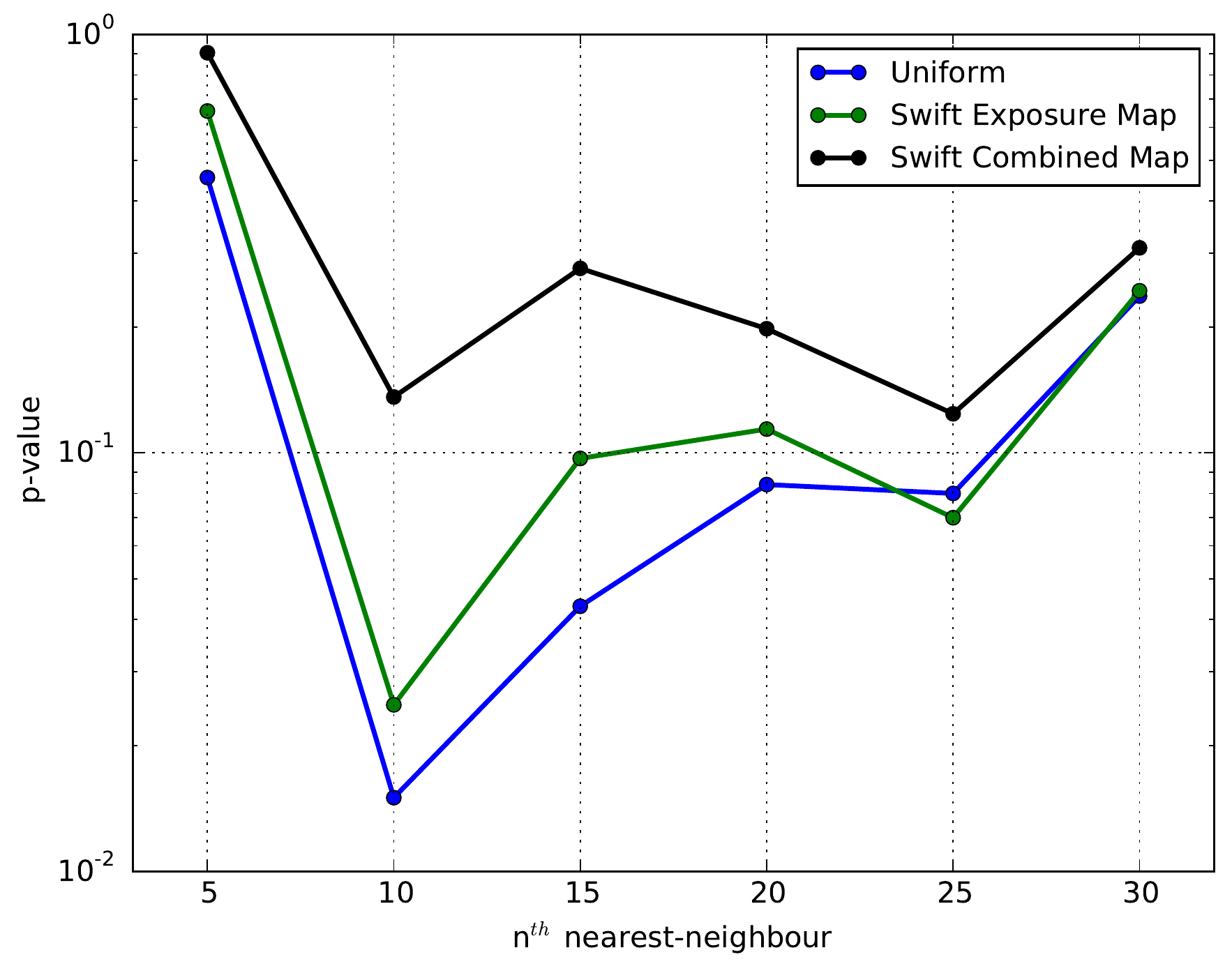}
\caption{Probability of observing the measured difference between the expected
distribution of the n-th nearest neighbor and the actual $Swift$ data for several values of n.
}\label{p_values_vs_nnnd}
\end{figure}

Another density indicator can be obtained from the cumulative distribution
of the n-th nearest-neighbor distance as was done by~\cite{Horvath2012}.
Figure~\ref{nnn_redshift} shows the cumulative distribution of the distance to the
10-th nearest neighbor in our sample of 34 GRBs with redshifts between 1.6 and 2.1
(shown in red). This is compared to the mean cumulative distribution of the
10-th nearest neighbor distance for GRBs distributed according to the $Swift$ exposure map
(green line), the $Swift$ map conditional on redshift measurement (black line)
and uniformly distributed GRBs (blue line). 
As before, 5000 samples were simulated with 34 GRBs per sample to calculate the reference cumulative 
distance distributions. The p-value was derived from the distribution of the maximum differences 
between the distribution of the 10-th nearest neighbor distance in each simulated sample and the
reference (mean) distribution for both scenarios under consideration.
The resulting p-value is 0.022 for simulated GRBs distributed according
to the $Swift$ exposure map, 0.127 for the combined $Swift$ map and 
0.016 for the uniform case. The results for other values of n
nearest-neighbors are shown in Figure~\ref{p_values_vs_nnnd}.

Probabilities from both analyses (based on density estimation and proximity) are consistent
with each other. However, we note that proximity analysis based on the combined $Swift$
prior seems to give a particularly large p-value.
Without any additional insight all probabilities obtained so far would indicate that observing
the actual sky distribution of the 34 GRBs in our sample is a somewhat unlikely
event. This is not sufficient, however, to conclude with good confidence
that there is a significant clustering of GRBs in the redshift range $1.6 < z < 2.1$.
Moreover, the density maps in other possible redshift bins look very flat and are
entirely consistent with random fluctuations. Perhaps the most serious problem
is that the quoted p-values do not take into account the (unknown) number of implicit trials
that occurred when the redshift range was selected from all possible slices of the
original data by \cite{Horvath2012}. Since there is a large overlap between our data set
and the one used by \cite{Horvath2012}, the corrected probabilities could easily be
an order of magnitude larger. Therefore, we find no significant evidence of redshift-dependent
clustering in the $Swift$ GRB sample.

\section{Duration Dependent Clustering of GRBs}\label{t90_clustering}

\subsection{GRB Sample}

\begin{figure}
\includegraphics[width=84mm]{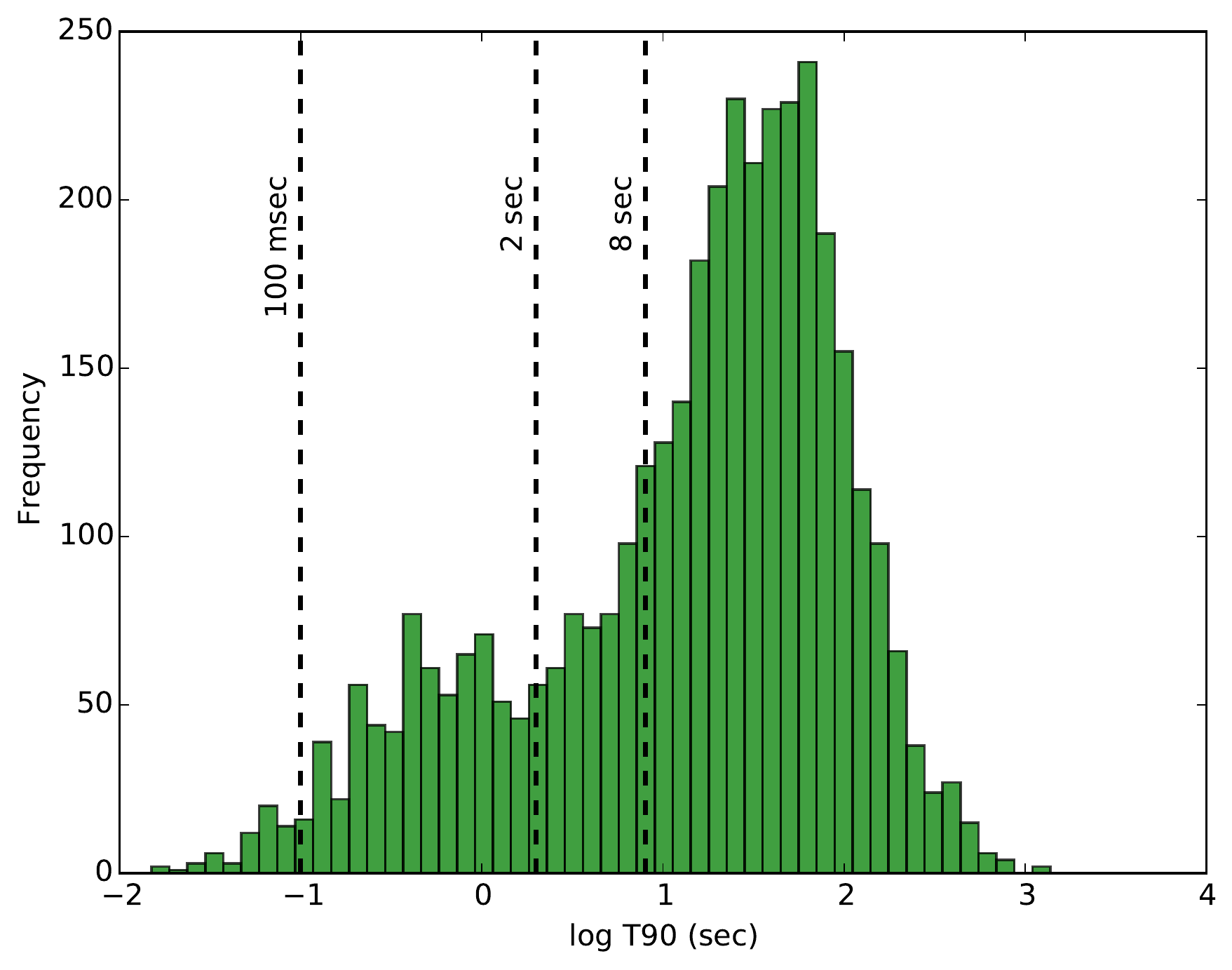}
\caption{Distribution of durations for our sample of 3798 bursts.}\label{t90histo}
\end{figure}

The two dominant populations of GRBs were originally distinguished based on
the bimodal distribution of burst durations: long GRBs with T90 $\ge$ 2 s
and short GRBs with T90 $<$ 2 s~\citep{Kouveliotou1993}. T90 is the time
interval that contains 90\% of the burst fluence centered on the mid point (i.e. starting at 5\%).
Several authors report evidence for additional GRB populations based on the distribution
of durations~\citep{Balazs1998, Cline1999, Meszaros2000, Litvin2001, Magliocchetti2003,
Cline2005, Cline2011}.
Duration-dependent clustering in the sky distribution of GRBs may help to identify
a new GRB population with distinct physical properties.
We investigated this possibility using several different samples of GRBs
detected by $Swift$, $Fermi$ Gamma-ray Burst Monitor (GBM) and CGRO's
(Compton Gamma Ray Observatory) Burst and Transient Source Experiment (BATSE).
Our data set includes 2037 BATSE GRBs
\footnote{\texttt{http://gammaray.msfc.nasa.gov/batse/grb/catalog/current/}}
~\citep{Paciesas1999}, 997 GBM GRBs~\citep{Gruber2014, Kienlin2014} and 889 $Swift$
GRBs~\citep{Lien2015}. Within this combined catalog, 125 bursts were detected by both
$Swift$ and GBM, and for those we used the $Swift$ observations because they provide
much better localizations. Our final combined sample includes a grand total of 3798 GRBs.
The corresponding T90 distribution is shown in Figure~\ref{t90histo}.
We used the same approach as in Section~\ref{z_clustering}
to investigate the duration-dependent clustering in the GRB sky distribution.

\subsection{Very Short Duration GRB Density Map}

\begin{figure*}
\includegraphics[width=0.9\textwidth]{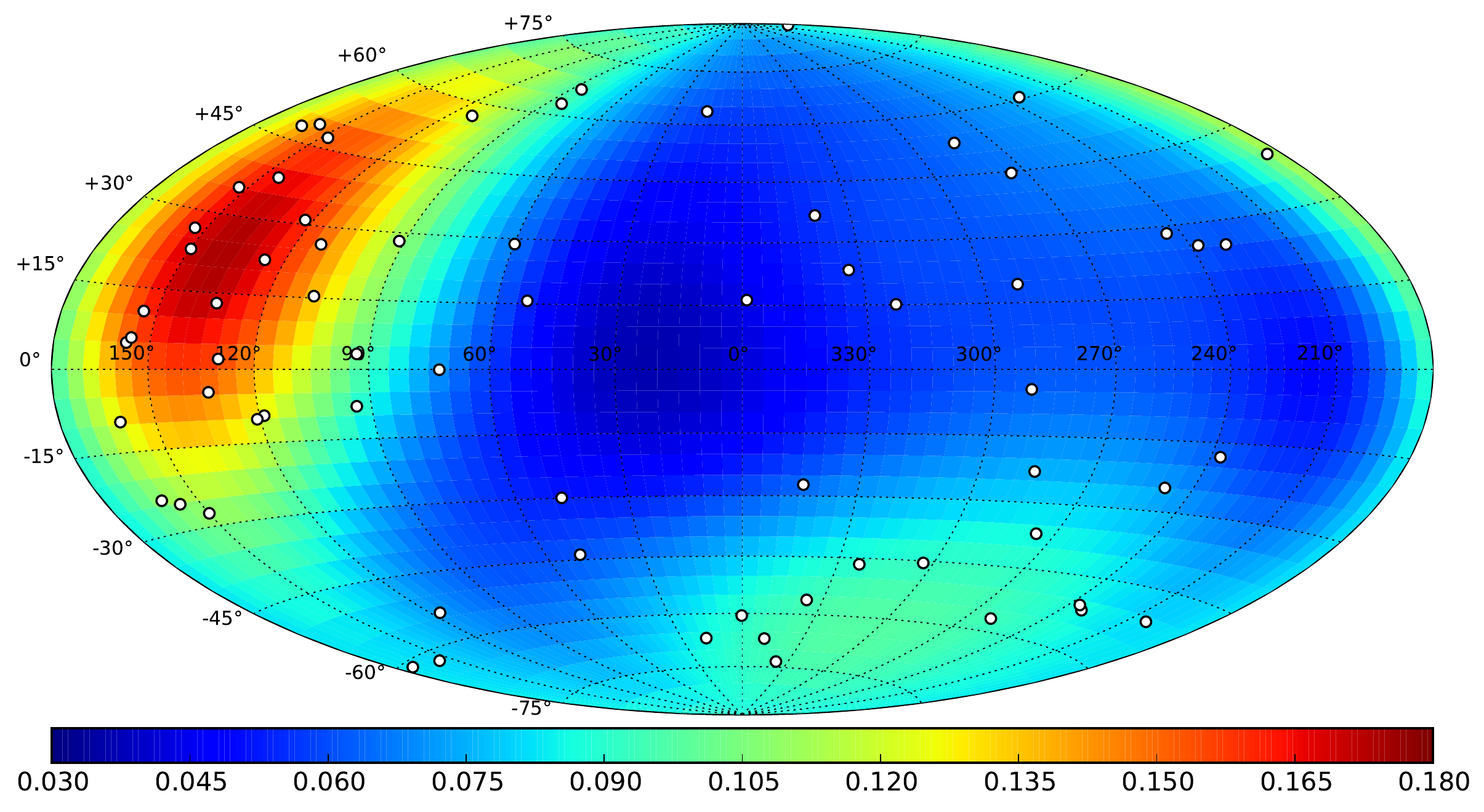}
\caption{Density map in galactic coordinates for a subsample of 68 very short
GRBs with T90 $\le$ 100 ms. The optimal smoothing length is 28 degrees.
Circles indicate the actual burst locations.
The maximum and the minimum density values in this map are 0.174 and 0.036,
correspondingly. The probability of generating this density contrast by chance
estimated using a Monte Carlo simulation is 0.0008 assuming that burst detections
follow the multi-instrument exposure map.}
\label{t90_100ms_density_map_duration}
\end{figure*}

There are 68 GRBs with T90 $\le$ 100 ms in our sample.
Figure~\ref{t90_100ms_density_map_duration} shows the corresponding density map
computed using the optimal smoothing length of 28 degrees.
A concentration of GRBs can be clearly seen toward the left side
of the map.
Using a different compilation of burst localizations Cline et al. found evidence
that very short GRBs tend to cluster in an area of the sky that roughly
coincides with the overdensity in Figure~\ref{t90_100ms_density_map_duration}
~\citep{Cline1999, Cline2005, Cline2011}. How significant is this density contrast?

\begin{figure}
\includegraphics[width=84mm]{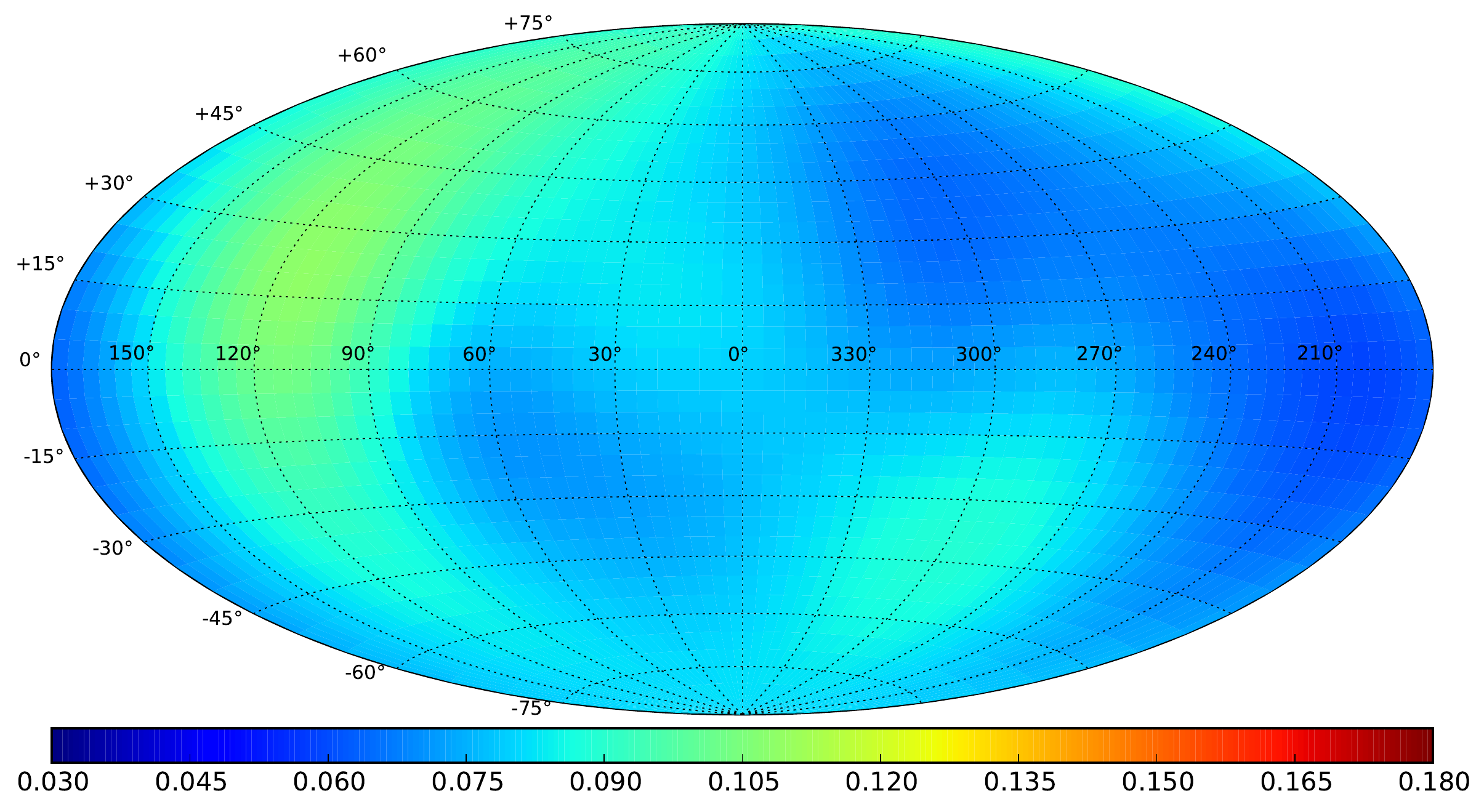}
\caption{$Swift$ exposure map derived assuming that long GRBs are uniformly distributed.
The map was computed using a sample of 808 long bursts and the optimal smoothing length
of 24 degrees.
}\label{expmap_swift}
\end{figure}

Unlike for $Swift$, exposure maps are not available for $Fermi$ and BATSE.
However, if we had a good sample of GRBs from a population for which the true
sky distribution is uniform, we could use the density map for that sample
as a proxy exposure map. We verified the validity of this approach
using long GRBs detected by $Swift$. The location and amplitude of the main features
in the resulting density map (Figure~\ref{expmap_swift}) are similar to the actual $Swift$
exposure map from Figure~\ref{expmap}. We therefore adopt this method to derive
exposure corrections for all GRB samples except those limited to bursts detected
by $Swift$. The correction is actually less important for multi-instrument
samples that tend to average out any local variations specific to a single instrument.
Although the p-values change slightly when a flat exposure map is assumed,
none of our conclusions depend on the correction.

\begin{figure}
\includegraphics[width=84mm]{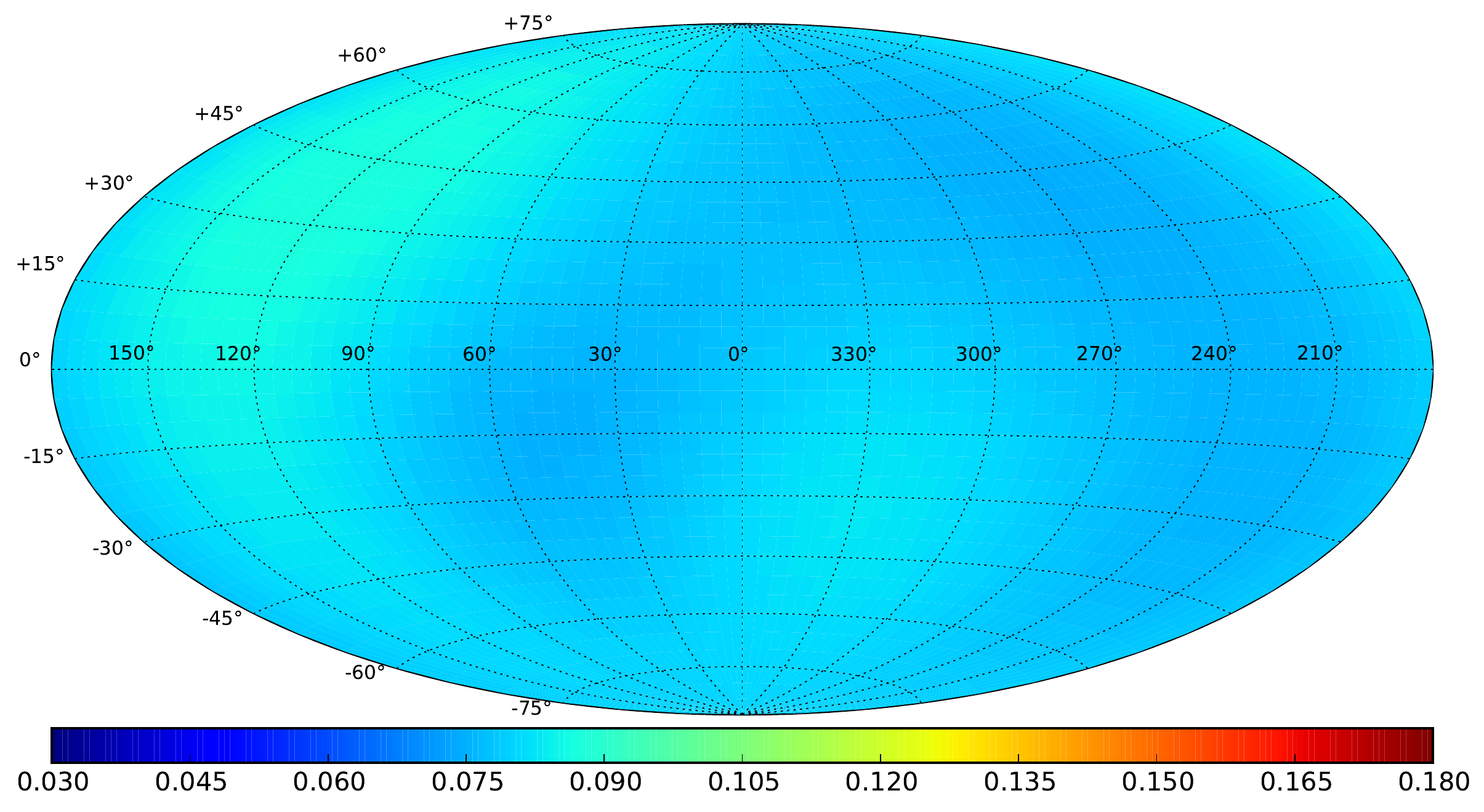}
\caption{Multi-instrument exposure map derived assuming that long GRBs are uniformly
distributed. The map was computed using a sample of 3063 long bursts detected by BATSE,
$Fermi$/GBM, and $Swift$/BAT and the optimal smoothing length of 31 degrees.
}\label{expmap_all}
\end{figure}

As before, we use Monte Carlo simulations
to estimate the significance of the anisotropy in the distribution of VSGRBs
(Figure~\ref{t90_100ms_density_map_duration}).
The effective exposure map for all three instruments computed assuming a uniform distribution
of long GRBs is shown in Figure~\ref{expmap_all}. The map is based on
3063 long GRBs detected by BATSE, $Fermi$/GBM and $Swift$/BAT. Note that the amplitude
variations of this map is only $\sim25$\%, much less that the factor of two variations
seen in the $Swift$ exposure map.
We draw 5000 samples of 68 GRBs with the probability density proportional to the exposure
map, compute density maps, and compare the resulting distribution of the peak density
with the observed value. The probability of generating
the observed density peak (Figure~\ref{t90_100ms_density_map_duration})
from a random fluctuation is 0.0008.

\begin{figure}
\includegraphics[width=84mm]{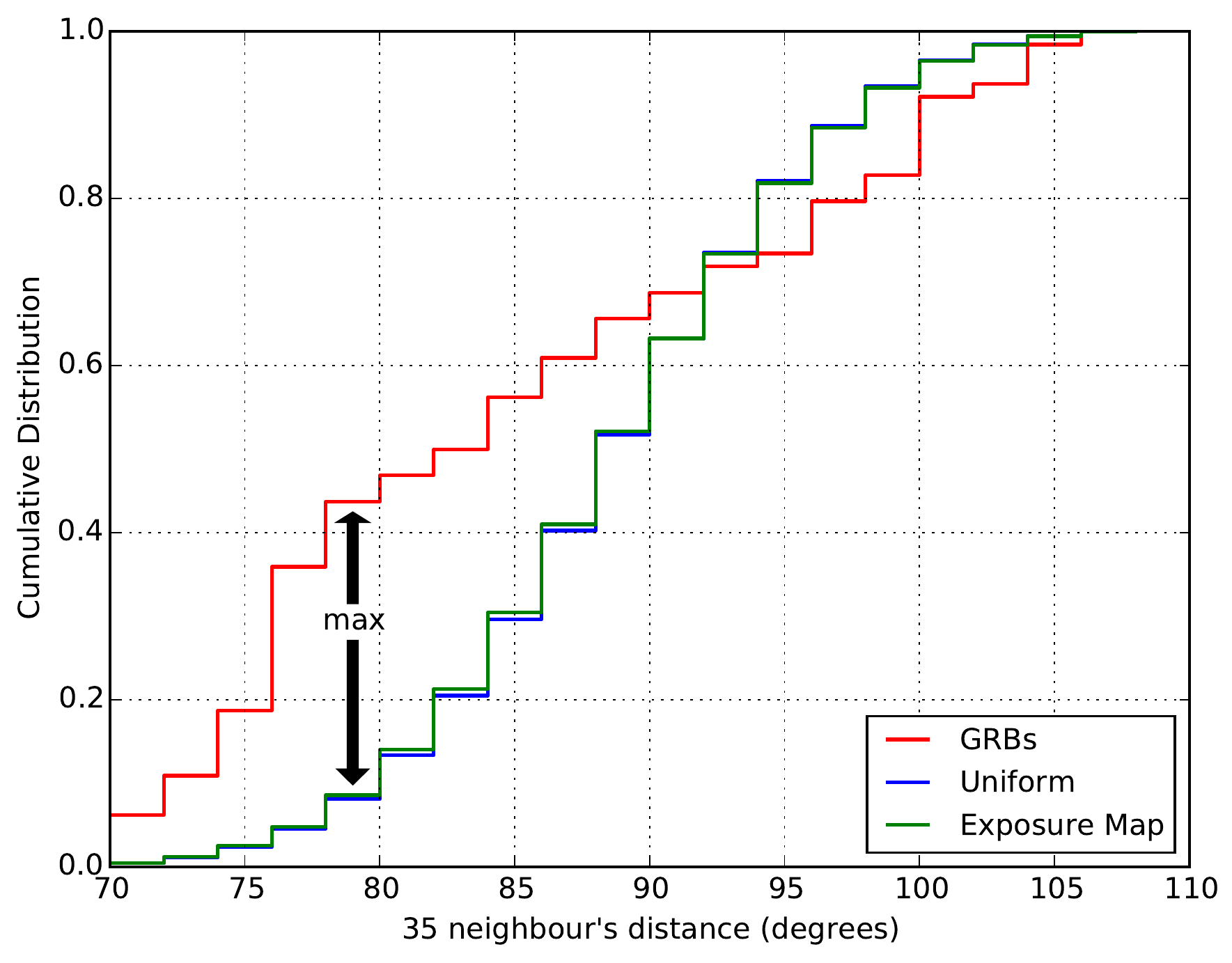}
\caption{Cumulative distribution of the 35-th nearest-neighbour distance
compared to the mean of 5000 simulated samples of 68 GRBs
(cf. Figure~\ref{t90_100ms_density_map_duration}).
The probability of the observed deviation from the uniform distribution
and the multi-instrument exposure map is, respectively, 0.0078 and 0.0084.
}\label{nnn_duration_t90}
\end{figure}

\begin{figure}
\includegraphics[width=84mm]{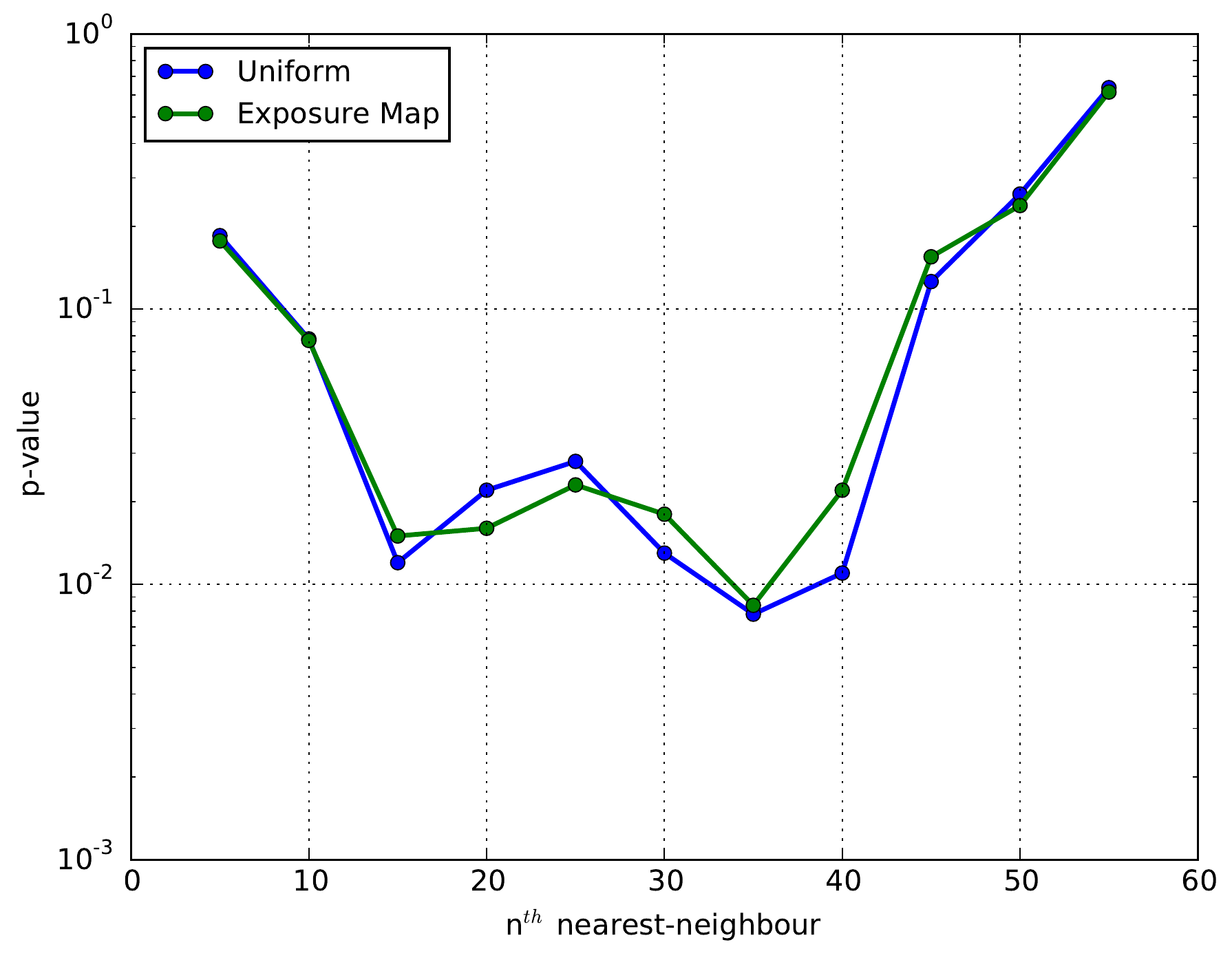}
\caption{Probability of the maximum deviation between the
observed distribution of the n-th nearest neighbour distance
and the mean of simulated samples of 68 VSGRBs.
The significance peaks for $n=35$ for both considered types of exposure maps.
}\label{p_values_vs_nnnd_duration}
\end{figure}

Similarly to Section~\ref{z_clustering}, we also examined the cumulative
n-th nearest-neighbor distance distributions for our sample of 68 VSGRBs.
Figure~\ref{nnn_duration_t90} shows the cumulative distance distribution
for the 35-th nearest neighbor (red) and the mean of 5000 samples drawn from a uniform
distribution (blue) and from the multi-instrument exposure map (green). As expected,
the two sets of simulated results are very similar because the multi-instrument exposure map
is very flat. The corresponding p-values are 0.0078 for the uniform distribution
and 0.0084 for the multi-instrument exposure map. Figure~\ref{p_values_vs_nnnd_duration}
shows how the p-values change as a function of the n-th nearest-neighbour.
The lowest p-value (highest significance) is reached for $n=35$ shown in
Figure~\ref{nnn_duration_t90}. Again, there is little difference between the
multi-instrument exposure map and the uniform distribution.

While the significance of the observed anisotropy in the
distribution of VSGRBs is not overwhelming, the probability of a chance alignment
is less than 1\%. Unfortunately, these probabilies could become much higher
when corrected for the unknown number of multiple trials incurred to select
the range of GRB durations for this clustering analysis.

\subsection{Short and Intermediate Duration GRB Density Map}

\begin{figure}
\includegraphics[width=84mm]{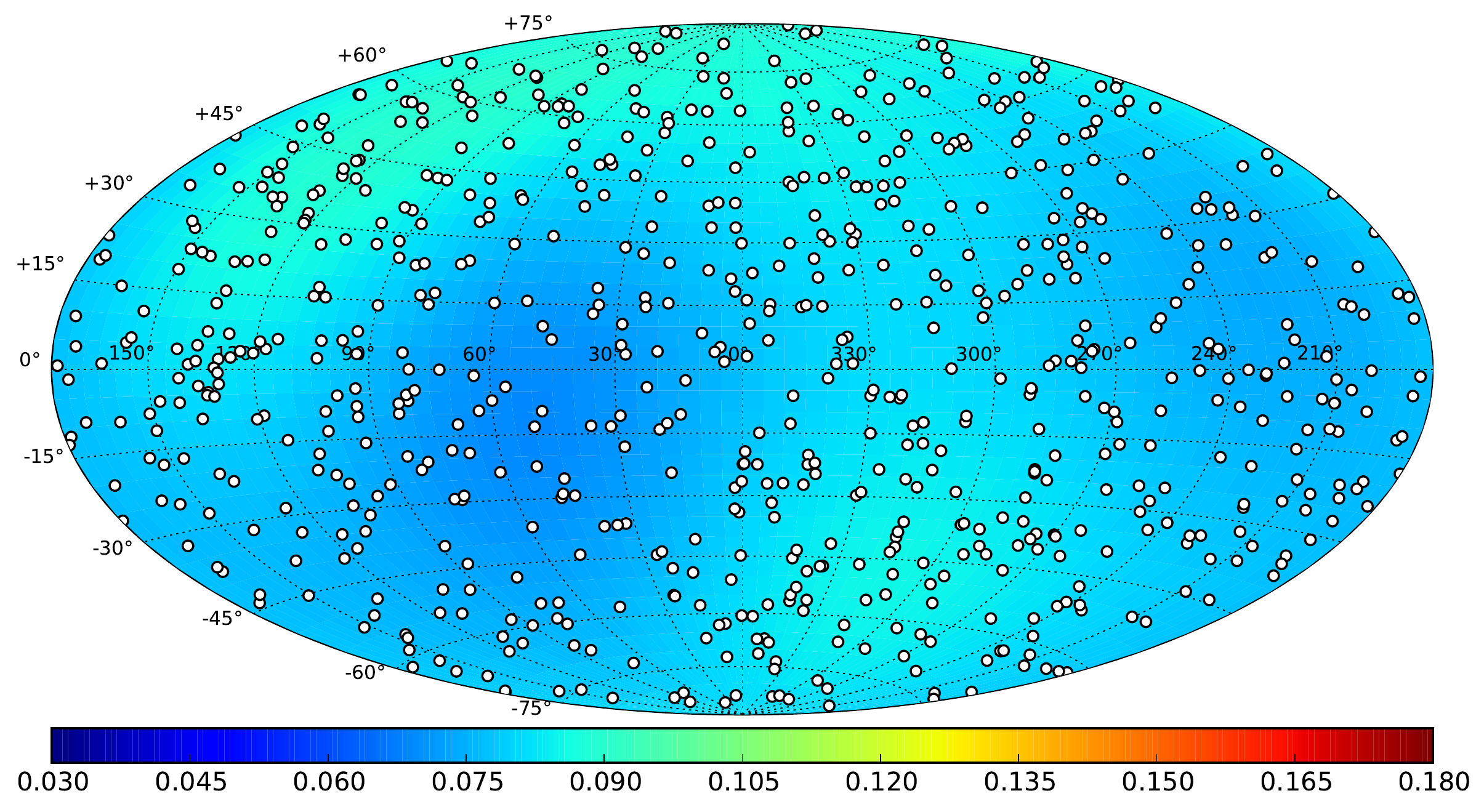}
\caption{Density map in galactic coordinates for a subsample of 735 short
GRBs with T90 $<$ 2 s. The optimal smoothing length is 35 degrees.
Circles indicate the actual burst locations.
The maximum and the minimum density values in this map are 0.09 and 0.07,
correspondingly.
The probability of generating this density contrast by chance estimated using a Monte Carlo
simulation is 0.3 assuming that burst detections follow
the multi-intrument exposure map.
}
\label{sgrb_density_map_duration}
\end{figure}

The short-duration bursts---thought to originate from compact binary mergers---are expected
to trace galaxies at cosmological distances. This
suggests that there should not be any significant anisotropies in their
sky distribution. Meanwhile \cite{Balazs1998} and \cite{Magliocchetti2003}
have reported evidence that supports the existence of such anisotropies.
We searched for clustering in our sample of 735 short GRBs defined here as having T90 $\le$ 2 s.
The resulting density map shown in Figure~\ref{sgrb_density_map_duration} is very flat
and allows little or no clustering. This visual impression is confirmed
by a high probability (p-value $\sim0.3$) of the observed density peak
from Monte Carlo simulations based on the multi-instrument exposure map.

\begin{figure}
\includegraphics[width=84mm]{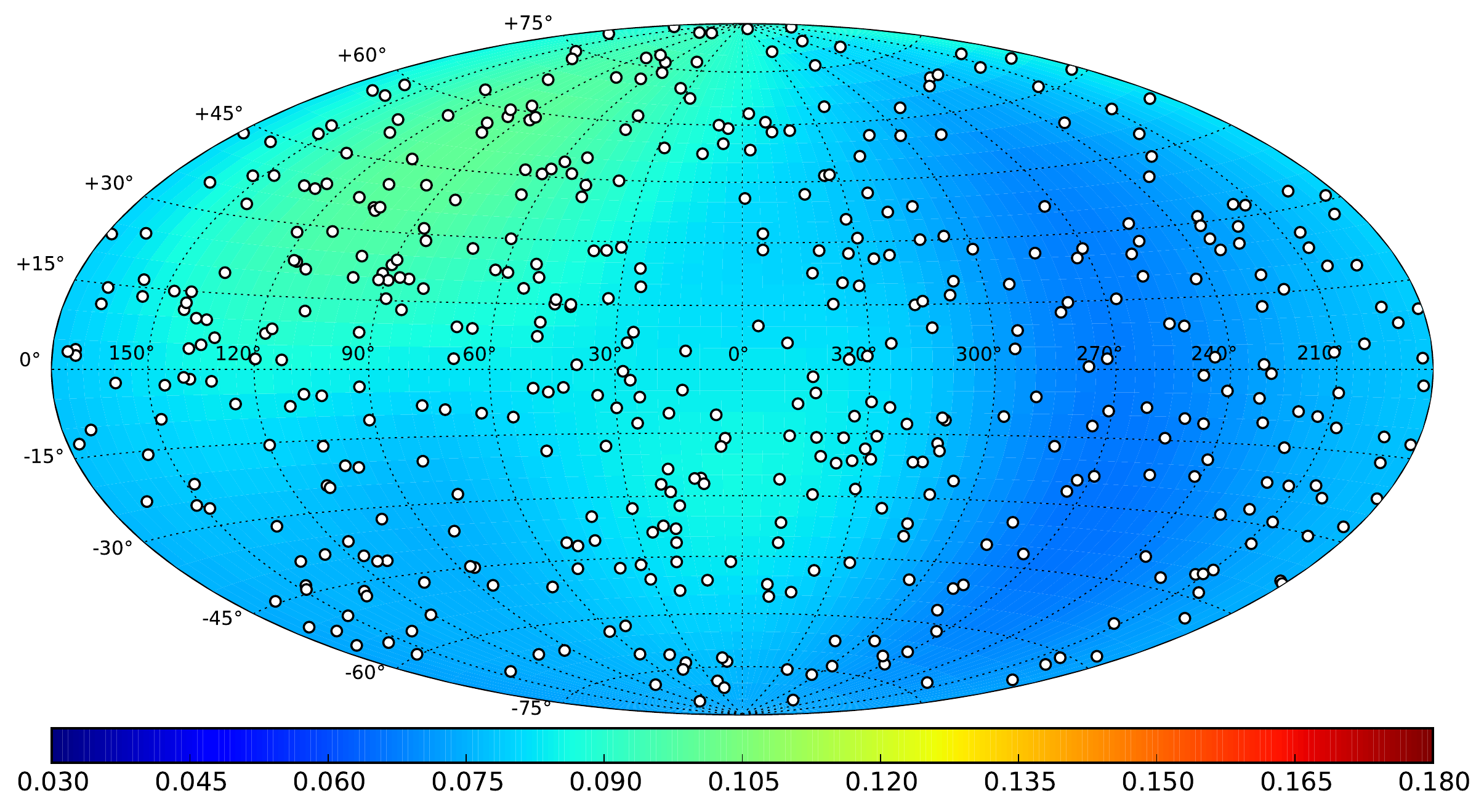}
\caption{Density map in galactic coordinates for a subsample of 468 short
GRBs with 2 s $<$ T90 $<$ 8 s. The optimal smoothing length is 32 degrees.
Circles indicate the actual burst locations.
The maximum and the minimum density values in this map are 0.1 and 0.07,
correspondingly.
The probability of generating this density contrast by chance estimated using a Monte Carlo
simulation is 0.05 assuming that burst detections follow
the multi-intrument exposure map.
}
\label{intgrb_density_map_duration}
\end{figure}

While the evidence that the intermediate-duration GRBs defined as having 2 s $<$ T90 $<$ 8 s
constitute a separate physical class is much weaker, there is some evidence that they
exhibit a noticeable level of anisotropy in their sky distribution~\citep{Meszaros2000,
Litvin2001}. Our data set includes 468 intermediate-duration GRBs.
Figure~\ref{intgrb_density_map_duration} shows the density map
for this sample with the optimal smoothing radius of 32 degrees.
We find no indication of significant clustering in the distribution of intermediate-duration
GRBs with the formal p-value of $\sim0.05$.

\section{Discussion}\label{discussion}

A recent analysis of GRB clustering across the redshift space
by \cite{Horvath2012} revealed a potential anisotropy in the redshift range
$1.6 < z < 2.1$. Using a sample of 31 GRBs detected by multiple instruments
these authors argued that there is evidence for a large scale structure
in this particular redshift bin. Our study based on 34 $Swift$ GRBs in the same
redshift range does not confirm the significance of features reported by \cite{Horvath2012}.
Although our estimated density maps show similar looking peaks and valleys
(Figure~\ref{z_density_map}), a rigorous Monte Carlo simulation demonstrates
that, given the small sample size, the observed density variations are consistent
with random fluctuations at a few percent probability level. This conclusion is reinforced when
we consider the effect of multiple trials necessary to identify one interesting redshift bin
from the entire GRB catalog. For example, 10 trials would roughly increase the probability
by an order of magnitude. Given the available data, the observed
clustering is not significant.

\begin{figure}
\includegraphics[width=84mm]{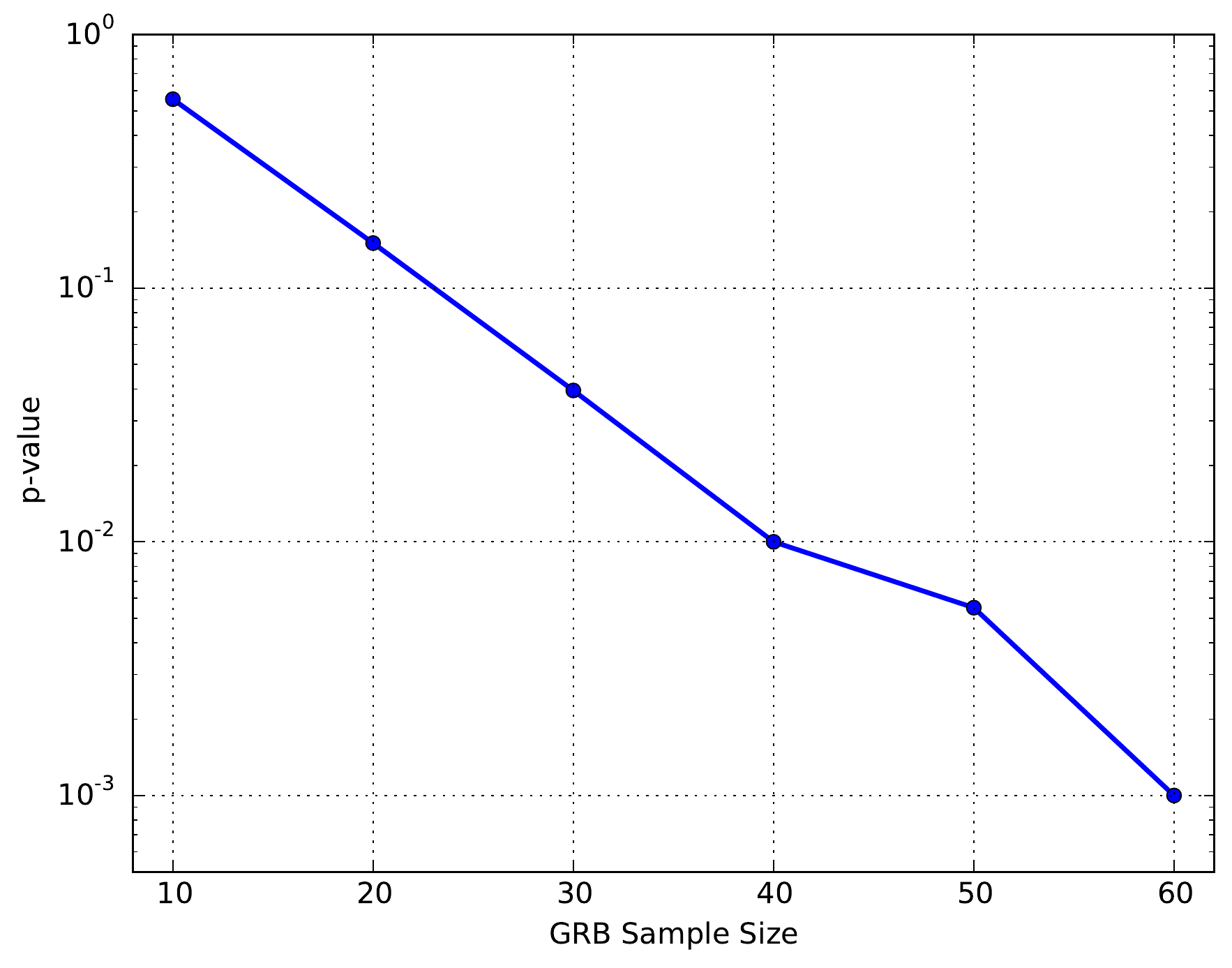}
\caption{Probability of observing a global density peak from Figure~\ref{z_density_map}
as a function of the sample size. The simulation is based on the $Swift$ exposure
map with 33 degree smoothing radius.}
\label{p_value_vs_sample_size}
\end{figure}

The probability of observing a purely random density fluctuation of a certain
amplitude clearly depends on the size of the sample. It is interesting
to investigate how many GRBs are needed in a redshift bin to exclude a chance alignment
with a reasonably good confidence (e.g. pre-trial p-value $\sim0.001$). We generated
samples of varying size using the $Swift$ exposure map and estimated the probability
of observing a global PDF maximum of 0.167 or higher corresponding to our actual measurement
from Figure~\ref{z_density_map}. The results are shown in Figure~\ref{p_value_vs_sample_size}.
More than 60 GRBs per redshift bin are required to reach the p-value of 0.001 in this case.
With the current rate of GRB discoveries and redshift measurements, it will likely take
longer than 10 years to accumulate a sufficiently large sample and make
a definitive statement about the presence of clustering, assuming that the currently
observed density contrast persists.

The most intriguing result from our duration-dependent clustering analysis
is the relatively strong anisotropy in the distribution of VSGRBs with T90 $\le$ 100 ms.
The pre-trial p-value of the density peak seen in Figure~\ref{t90_100ms_density_map_duration}
is 0.0008. Somewhat lower confidence levels were obtained from the analysis
of the cumulative n-th nearest-neighbor distance distribution.
It is not clear in this case how many trials were incurred when this particular
range of durations was selected. In our analysis we examined four duration bins:
T90 $\le$ 100 ms, T90 $\le$ 2 s, 2 s $<$ T90 $<$ 8 s and T90 $>$ 2 s. However,
in order to properly account for trials, one needs to also consider the ones
incurred by previous authors in their analysis. Ultimately, this unknown number of trials
will degrade the significance of the clustering seen in the distribution of VSGRBs.
If confirmed by future GRB detections, this clustering may potentially
help to identify a new population of GRBs.

\cite{Cline2011} proposes that VSGRBs may originate from evaporating primordial black holes
(PBH) in the solar neighborhood. The main problem with this scenario, assuming that there is
in fact a significant clustering, is that the PBH sources would have to reside preferentially
in a relatively confined region of the solar neighborhood consistent with the Galactic anti-center.
Another problem is that the typical time profile of these bursts exhibits relatively
complex structure with multiple peaks \citep{Czerny2011}. This characteristic is
inconsistent with the final stages of the black hole evaporation, when the predicted emission
from the black hole is smooth as it only depends on one parameter, the
mass~\citep{Carr2010,Ukwatta2015icrc1}. All PBH bursts are therefore expected to look
similar and have a single peak. The observed variety of VSGRB light curves argues
against the PBH burst origin for most of the population.

The fact that some VSGRBs may be at cosmological distances is also incompatible with the PBH
hypothesis. The initial mass of PBHs expiring today is $\sim 5.0 \times 10^{11}$ kg
~\citep{Carr2010,Ukwatta2015icrc1}.
During the final stages of the burst, only about $\sim 10^{5}$ kg is left in a typical PBH.
Even assuming that all mass is converted into photons in the keV/MeV energy range,
the maximum possible distance for the PBH to remain detectable is less than few parsecs.
In our sample there are 13 $Swift$ GRBs in the VSGRB category. Three of those coincide
with host galaxies at known redshifts: GRB 050509B, GRB 060502B, and GRB 100206A.
While the host galaxy redshift measurement alone does not rule out the PBH origin of some
VSGRBs because of the possibility of chance associations, the presence of the long lived
lower energy emission does. One of the three bursts in question (GRB 050509B) displayed
an X-ray afterglow\citep{CastroTirado2005} that excludes the PBH origin for this particular
burst. However, it is still possible that some VSGRBs are in fact due to PBH bursts.

Other possible progenitors for VSGRBs are mergers of binary systems.
This scenario is an extension of the standard model for short GRBs.
It is possible to tweak merger models to generate time scales
observed in VSGRBs~\citep{Czerny2011}. However, in this case it is still
difficult to account for the observed potential clustering of VSGRBs.

\section{Conclusions}\label{conclusions}

We analysed the redshift- and duration-dependent clustering in the sky distribution
of GRBs using the Gaussian kernel density estimator and the n-th nearest neighbor distance.
Contrary to previous reports \citep{Horvath2012}, our redshift-dependent analysis did not
provide evidence of significant clustering. This is especially the case after considering
the number of trials incurred to find a hypothetical overdensity that requires
repeated redshift binning and multiple visual searches of density maps.
The duration-dependent analysis demonstrated that neither short duration
nor intermediate duration GRBs display any significant clustering.
However, very short duration GRBs appear to show some degree of clustering.
This has been previously noted by \cite{Cline1999,Cline2005} as well as \cite{Cline2011}
and warrants further attention as it may ultimately provide evidence for the existence
of a separate source population.

The apparent angular distribution of long GRBs detected by $Swift$ is proportional
(within the statistical error) to the total exposure time for a given line of sight.
In other words, the true sky distribution of long GRBs is very uniform and
can be used to construct approximate exposure maps for various instruments.
This technique will be particularly useful for multi-instrument exposure maps
and in situations where it is not straightforward to calculate the exposure map directly.
A good example of the latter is the Inter-Planetary Network (IPN).

\section*{Acknowledgments}

This work was funded by the US Department of Energy. TNU acknowledges support
from the Laboratory Directed Research and Development program at 
the Los Alamos National Laboratory.
We thank Brenda Dingus, Pat Harding, Krista Smith and Kevin Hurley
for useful conversations on the analysis. We also thank the referee
Jean-Luc Atteia for comments that significantly improved the paper. 

\bibliographystyle{plainnat}
\bibliography{my_references}


\bsp

\label{lastpage}

\end{document}